\documentclass[aps,superscriptaddress,showpacs,amsmath,amssymb,longbibliography,twocolumn]{revtex4-1}
\usepackage[T1]{fontenc} 
\usepackage{color} 
\usepackage[english]{babel} 
\usepackage{verbatim} 
\usepackage{float} 
\usepackage{graphicx} 
\usepackage[pdftex, colorlinks=true, linkcolor=blue, citecolor=blue,urlcolor=blue]{hyperref}

\definecolor{smaragd}{cmyk}{.9,0.3,.9,.1}

 \pdfpageheight\paperheight \pdfpagewidth\paperwidth

 \usepackage{pdfcomment} 
\usepackage{environ} 
\RenewEnviron{comment}{\pdfcomment{\BODY}}

\newcommand{\en}{{\varepsilon_{n\textbf{k}}}} 
\newcommand{\enp}{{\varepsilon_{n'\textbf{k}'}}}

\makeatother

\begin{document}

\title{Photocarrier thermalization bottleneck in graphene}

\author{Dinesh Yadav}
\affiliation{Okinawa Institute of Science and Technology Graduate University, Onna-son, Okinawa 904-0395, Japan}
\affiliation{Department of Physics, University of Konstanz, D-78457 Konstanz,
  Germany}
\author{Maxim Trushin}
\affiliation{Centre for Advanced 2D Materials, National University of Singapore, 6 Science Drive 2, Singapore 117546}
\author{Fabian Pauly}
\affiliation{Okinawa Institute of Science and Technology Graduate University, Onna-son, Okinawa 904-0395, Japan}
\affiliation{Department of Physics, University of Konstanz, D-78457 Konstanz,
  Germany}

\date{\today} 

\begin{abstract}
We present an ab-initio study of photocarrier dynamics in graphene due to
electron-phonon (EP) interactions.  Using the Boltzmann relaxation-time
approximation with parameters determined from density functional theory (DFT)
and a complementary, explicitly solvable model we show that the photocarrier
thermalization time changes by orders of magnitude, when the excitation energy
is reduced from 1~eV to the 100 meV range. In detail, the ultrafast
thermalization at low temperatures takes place on a femtosecond timescale via
optical phonon emission, but slows down to picoseconds once excitation
energies become comparable with these optical phonon energy
quanta. In the latter regime, thermalization times exhibit a pronounced
dependence on temperature. Our DFT model includes all the inter- and intraband
transitions due to EP scattering. Thanks to the high melting point of
graphene we extend our studies up to 2000~K and show that such high
temperatures reduce the photocarrier thermalization time through phonon
absorption.
\end{abstract}

\maketitle

\section{Introduction}
Recent progress in nanotechnology has made it possible to fabricate
high-quality materials that are only one atom thick and hence reach the
fundamental two-dimensional (2D) limit for solid crystals
\cite{cao2015quality}. Due to their ultimate thinness these materials
demonstrate various properties that are qualitatively different from those of
the three-dimensional parent crystals and, at the same time, are found to be
useful in photodetection and photovoltaic applications
\cite{Nanoscale2015roadmap}. Indeed, the central phenomenon employed in
photodetection and photovoltaics is the conversion of light energy into
electricity. It is a quantum conversion process, employing absorption of
photons to deliver photoexcited carriers to an external circuit, where they do
electrical work \cite{Nelson2004}. There are two obvious strategies for
increasing the amount of energy transferred by photocarriers. One can try to
speed up the photocarrier extraction such that the carriers are collected,
while they are still hot or even out of thermal equilibrium. Alternatively,
one can try to slow down the cooling or photocarrier thermalization for the
same purpose.  

Graphene in a combination with other 2D semiconductors offers an interesting
opportunity to employ both strategies.  Thanks to the extremely small
thickness of the junctions between 2D materials (also known as van der Waals
heterostructures \cite{Science2016novo}), interlayer photocarrier transport
may occur faster than the intralayer relaxation processes
\cite{NatPhys2016ma}. At the same time the optical phonon emission is strongly
suppressed for low-energy excitations in graphene due to unusually high energy
quanta of optical phonons \cite{mihnev2016microscopic,cooling_in_graphene_PhysRevLett.117.087401}. As a
consequence, the photocarriers can be extracted well before they thermalize
and dissipate useful energy by means of phonon emission.  By incorporating
graphene into a heterostructure, we can combine the two strategies in one
optoelectronic device. In this way the photoresponse can be substantially
increased simultaneously to the device performance.  In this paper, we focus
on the photocarrier evolution in graphene, providing conclusive evidence for
the existence of a thermalization bottleneck that makes such applications
possible.

The photocarrier dynamics in graphene has been studied experimentally by means
of pump-probe spectroscopy as well as time- and angle-resolved photoemission
spectroscopy
\citep{Daniela_Nature_communication_Brida2013,gr_pp1_PhysRevB.83.153410,gr_pp2_PhysRevLett.105.127404,gr_pp3_doi:10.1021/nn200419z,gr_pp4,Direct_view_of_hot_carrier_in_Graphene_PhysRevLett.111.027403,gierz2017probing,aeschlimann2017ultrafast, Trushin_pseudospin}.
In the experiments the photoexcited carriers lie far above the Dirac point (by
more than $1$~eV), and the ultrafast relaxation of hot carriers is mainly
attributed to optical phonon emission and carrier-carrier scattering, taking
place within 150-170~fs
\citep{Direct_view_of_hot_carrier_in_Graphene_PhysRevLett.111.027403,Daniela_Nature_communication_Brida2013}.
Excitations below the highest optical phonon energy (of around $200$~meV
in graphene) have been studied in
Refs.~\citep{Acoustic_graphene_PhysRevLett.107.237401,cooling_in_graphene_PhysRevLett.117.087401,mihnev2016microscopic},
where it has been observed that the relaxation time is drastically enhanced
from the femtosecond to the picosecond timescale.  Despite multiple
theoretical contributions in the field of photocarrier thermalization and
cooling in graphene \citep{mihnev2016microscopic, PRB2009kubakaddi,
  PRL2009bistritzer, PRB2009tse, PRB2012low, PRB2011Malic, PRB2011kim,
  PRL2012song, Malic2012efficient, tomadin2013nonequilibrium,
  menabde2017interface}, the leading role of phonons in this enhancement still
requires conclusive evidence from a parameter-free ab-initio point of view. 

In what follows, we present an ab-initio approach to calculate the relaxation
time of photoexcited carriers in graphene, relying on EP scattering.  We
use DFT to calculate EP scattering rates. Inclusion of contributions
arising from all the optical and acoustical phonon branches in the whole
Brillouin zone (BZ) makes it possible to calculate the energy-dependent
relaxation time without adjustable parameters.  Moreover, we include inter-
and intraband processes, arising from the EP scattering. We investigate
the relaxation time for different excitation energies from 0.05 to 0.8~eV and,
due to the high melting point of graphene at around 5000~K
\cite{melting_graphene}, over a wide range of temperatures from 0 to
2000~K. Finally, we develop an explicitly solvable model to understand the
energy dependence of the photocarrier thermalization.

Our paper is organized as follows. In Sec.~\ref{sec:theory} we present the
theoretical approaches used in this work. Next, we discuss the results
obtained within the models in Sec.~\ref{sec:results} before we end with a
summary and outlook in Sec.~\ref{sec:summary-outlook}.

\section{Theoretical approaches}\label{sec:theory}
In this section, we describe the theoretical approaches that we apply. In
subsection~\ref{subsec:method-ab-initio} these are the details of our DFT
calculations to determine electronic and phononic properties. Subsequently, we
present in subsection~\ref{subsec:method-time-evolution} the Boltzmann
equations in the relaxation-time approximation, as employed to determine the
photocarrier dynamics. In subsection \ref{subsec:method-analytical} we finally
discuss simplifications to the relaxation-time approximation in order to
obtain an explicitly solvable model.

\subsection{Ab-initio theory for electronic and phononic properties}\label{subsec:method-ab-initio}

We use DFT within the local density approximation (LDA) to calculate the
ground-state electronic properties of graphene with \textsc{Quantum Espresso}
\citep{QUANTUMESPRESSO_0953-8984-21-39-395502}. We employ a plane-wave basis
set with a kinetic energy cutoff of 110~Ry, a charge density cutoff of 440~Ry
and a Troullier-Martins pseudopotential for carbon with a $2s^{2}2p^{2}$
valence configuration \citep{TM_pseudopotential_PhysRevB.43.1993}. The unit
cell of graphene is relaxed with the help of the
Broyden-Fletcher-Goldfarb-Shanno algorithm until the net force on atoms is
less than $10^{-6}$~Ry/a.u., and total energy changes are below
$10^{-8}$~Ry. A vacuum of 20~\textup{\AA} along the out-of-plane direction is
used to avoid artificial interactions with periodic images of the graphene
sheet, and the BZ is sampled with a $45\times45\times1$
$\Gamma$-centered $\mathbf{k}$-grid. We construct Wannier functions to get
localized orbitals from plane-wave eigenfunctions. By interpolating
wavefunctions, we finally obtain electronic eigenenergies, dynamical matrices
and EP couplings on fine grids in the BZ
\citep{Wannierorbitals}. We calculate the phonon dispersion spectrum of
graphene through density functional perturbation theory (DFPT)
\citep{DFPT_RevModPhys.73.515}, employing a $12\times12\times1$
$\mathbf{q}$-grid to evaluate phonon dynamical matrices.

\begin{figure}[!tb] \centering{}\includegraphics[width=1.0\columnwidth]{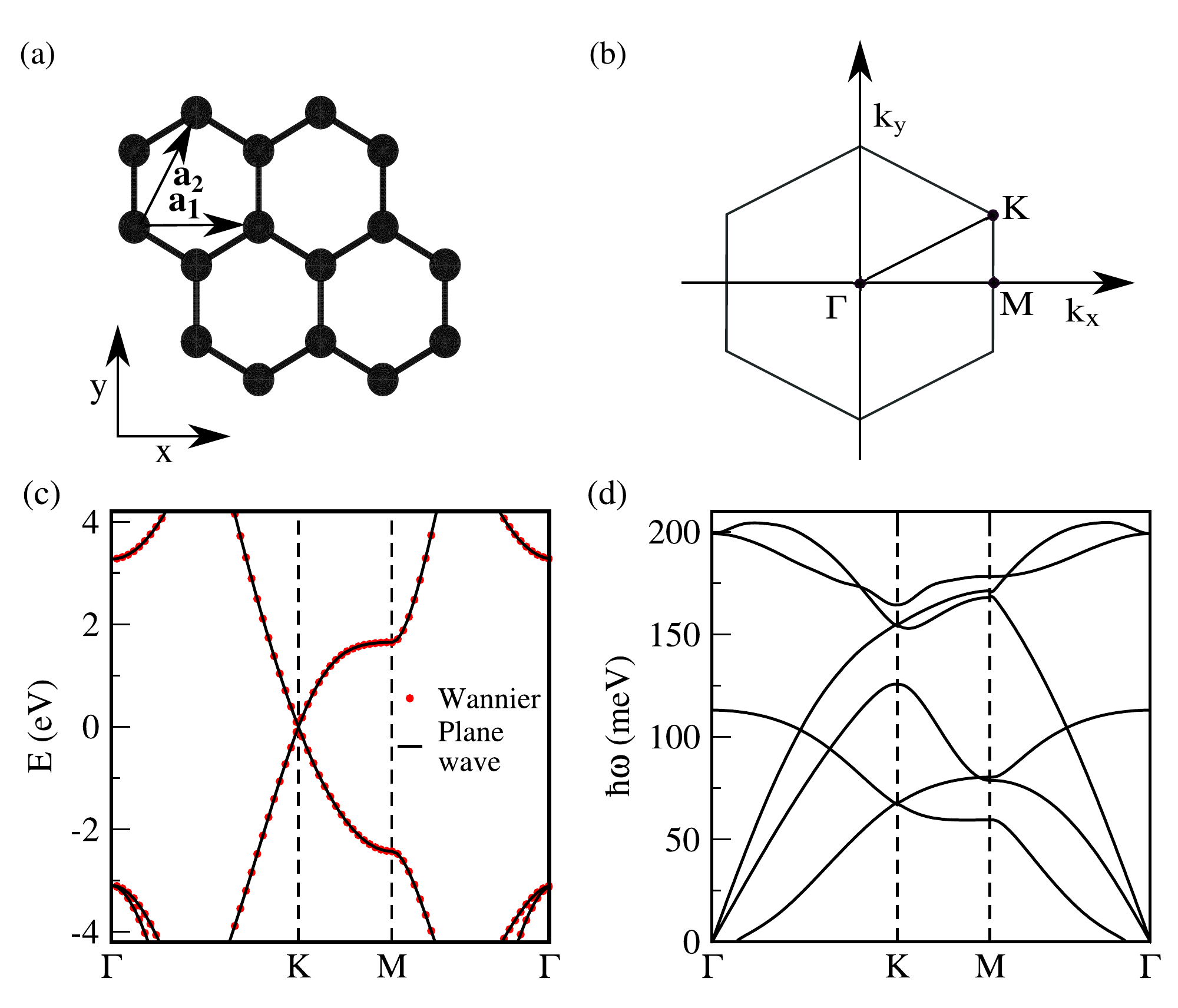}
  \caption{(a) Lattice structure of graphene with in-plane lattice vectors
      $\mathbf{a}_1$ and $\mathbf{a}_2$. (b) Reciprocal lattice of graphene
      with high symmetry points in the first BZ. (c) Electronic band
    structure of graphene, as obtained directly from the calculations with the
    plane-wave basis set and the corresponding curve from the Wannier-function
    formalism. (d) Phonon band structure of graphene.} \label{fig:bandstruct}
\end{figure}

By performing the DFT procedures, we obtain an optimized in-plane lattice
constant of graphene of $a=|\mathbf{a}_1|$ = $|\mathbf{a}_2|$ =
2.436~\textup{\AA}, see Fig.~\ref{fig:bandstruct}(a), which is in good
agreement with previous reports of 2.458~\textup{\AA}
\citep{graphene_phonon_DFT_PhysRevB.77.125401}. We calculate electronic
  and phononic band structures along high symmetry lines of the first BZ, as
  plotted in Fig.~\ref{fig:bandstruct}(b). Fig.~\ref{fig:bandstruct}(c) shows
the electronic band structure, as computed from DFT with plane waves. The
excellent agreement with those determined through the Wannier function method
demonstrates the high quality of the interpolated localized orbitals. The
phonon dispersion is finally displayed in
Fig.~\ref{fig:bandstruct}(d). Longitudinal optical and transverse optical
phonon modes of graphene at the $\Gamma$-point are degenerate at an energy of
198.37~meV, which matches well with a previously reported value of
197.75~meV \citep{graphene_phonon_DFT_PhysRevB.77.125401}.

Having determined electronic and phononic band structures, we calculate the
electronic self-energy $\Sigma_{n\mathbf{k}}(T)$ due to the EP interaction
for the electronic eigenstate $|n\mathbf{k}\rangle$ with the \textsc{EPW}
code. It is defined as follows \citep{EPW_package_PONCE2016116}
\begin{widetext} 
  \begin{multline} 
    \Sigma_{n\mathbf{k}}(T)=\sum_{m,p}\int_{\text{BZ}}\frac{d^3q}{\Omega_{\text{BZ}}}\vert
    g_{mn,p}(\mathbf{k},\mathbf{q})\vert^{2}\times\left[\frac{N_{\hbar\omega_{p\mathbf{q}}}(T)+f^{(0)}_{\varepsilon_{m\mathbf{k+q}}}(T)}{\en-(\varepsilon_{m\mathbf{k+q}}-\varepsilon_{\text{F}})+\hbar\omega_{p\mathbf{q}}+\text{i}\eta}+\frac{N_{\hbar\omega_{p\mathbf{q}}}(T)+1-f^{(0)}_{\varepsilon_{m\mathbf{k+q}}}(T)}{\varepsilon_{n\mathbf{k}}-(\varepsilon_{m\mathbf{k+q}}-\varepsilon_{\text{F}})-\hbar\omega_{p\mathbf{q}}+\text{i}\eta}\right],\label{eq:self-energy}
  \end{multline}
\end{widetext}
where $n$ is the band index, $\mathbf{k}$ is an electronic wave vector in the
BZ, $\hbar\omega_{p\mathbf{q}}$ is the energy of the phonon of branch $p$ at
wave vector $\mathbf{q}$, $\varepsilon_{\text{F}}=0$ is the Fermi energy,
$f^{(0)}_{\varepsilon_{n\mathbf{k}}}(T)=1/[\exp(\frac{\varepsilon_{n\textbf{k}}-\varepsilon_{\text{F}}}{k_{\text{B}}T})+1]$
is the Fermi-Dirac distribution,
$N_{\hbar\omega_{p\mathbf{q}}}(T)=1/[\exp(\frac{\hbar\omega_{p\mathbf{q}}}{k_{\text{B}}T})-1]$
is the Bose function, $\Omega_{\text{BZ}}$ is the volume of the BZ, and
$\eta=10~\text{meV}$ is the small broadening parameter. The EP matrix elements
are defined as \citep{EPW_package_PONCE2016116}
\begin{equation}
  g_{mn,p}(\mathbf{k,q})=\frac{1}{\sqrt{2\omega_{p\mathbf{q}}}}\left<m\mathbf{k+q}|\partial_{p\mathbf{q}}V|n\mathbf{k}\right> 
\end{equation}
and provide information about the scattering processes happening between the
Kohn-Sham states $|m\textbf{k+q}\rangle$ and $|n\textbf{k}\rangle$, as
mediated by the derivative $\partial_{p\mathbf{q}}V$ of the self-consistent
Kohn-Sham potential with respect to the phonon wavevector \textbf{q} in branch
$p$. Note that we assume that electron and phonon baths are at same
temperature $T$. The first term in the brackets of Eq.~(\ref{eq:self-energy})
can be seen as arising from absorption of phonons and the second one from
their emission.

To obtain converged results for Eq.~(\ref{eq:self-energy}), we first calculate
the electronic and vibrational states on a $36\times36\times1$ \textbf{k}-grid
and a $12\times12\times1$ \textbf{q}-grid using DFT and DFPT with plane-wave
basis functions \citep{QUANTUMESPRESSO_0953-8984-21-39-395502},
respectively. Finally, the electron eigenenergies, wavefunctions and phonon
dynamical matrices are interpolated on fine grids using Wannier functions
\citep{Wannier90_MOSTOFI20142309}. We use a $1200\times1200\times1$
\textbf{k}-grid and a $300\times300\times1$ \textbf{q}-grid, which we find
necessary to accurately map out the whole BZ and to converge the
integral over \textbf{q} in Eq.~(\ref{eq:self-energy}).

We assume that electronic wavefunctions and
phonon dynamical matrixes do not change with EP interactions
\citep{EPW_package_PONCE2016116}. The EP scattering time, resolved according
to the electronic band and momentum, is calculated as
\begin{equation}
  \tau_{n\mathbf{k}}(T)=\frac{\hbar}{2\text{Im}[\Sigma_{n\textbf{k}}(T)]}\label{eq:tau-nk}
\end{equation}
from Eq.~(\ref{eq:self-energy}) by using the imaginary part of the
self-energy.

\subsection{Time-evolution of excited charge carriers}\label{subsec:method-time-evolution}

The time evolution of the electronic occupation $\tilde{f}_{n\textbf{k}}(t,T)$
is calculated using the Boltzmann equation in the relaxation-time
approximation
\begin{equation}
  \frac{d\tilde{f}_{n\textbf{k}}(t,T)}{dt}=-\frac{\tilde{f}_{n\textbf{k}}(t,T)-f^{(0)}_{\varepsilon_{n\textbf{k}}}(T)}{\tau_{n\textbf{k}}}\label{eq:Boltzmann}
\end{equation}
with the solution
\begin{equation}
    \tilde{f}_{n\textbf{k}}(t,T)=f^{(0)}_{\varepsilon_{n\textbf{k}}}(T)+e^{-\frac{t}{\tau_{n\textbf{k}}}}[\tilde{f}_{n\textbf{k}}(0,T)-f^{(0)}_{\varepsilon_n\textbf{k}}(T)],\label{eq:Boltzmann-solution}
\end{equation}
if the excitation is assumed to happen at time $t=0$.
Eq.~(\ref{eq:Boltzmann-solution}) states that when the system is weakly
perturbed, the perturbation decays exponentially with the scattering time
$\tau_{n\textbf{k}}$ to restore the equilibrium Fermi-Dirac distribution
$f^{(0)}_{\varepsilon_{n\textbf{k}}}(T)$
\citep{Lundstrom_Book_lundstrom_2000}. The tilde sign indicates the time
dependence of the occupation function.

We generate the initial hot-carrier occupation $\tilde{f}_{n\textbf{k}}(0,T)$
as a combination of a Fermi-Dirac distribution $f^{(0)}_{\varepsilon_{n\textbf{k}}}(T)$ at
the temperature $T$ and a Gaussian peak at energy $+\zeta$ for
electrons in the conduction band
($\varepsilon_{n\textbf{k}}>\varepsilon_{\text{F}}$) and $-\zeta$ for the
holes in the valence band ($\varepsilon_{n\textbf{k}}<\varepsilon_{\text{F}}$)
as
\begin{equation} 
  \tilde{f}_{n\textbf{k}}(0,T)=f^{(0)}_{\varepsilon_{n\textbf{k}}}(T)\begin{cases}+\frac{\lambda_{\text{e}}}{\sqrt{2\pi\sigma^{2}}}e^{\frac{(\varepsilon_{n\textbf{k}}-\zeta)^{2}}{2\sigma^{2}}},
  & \varepsilon_{n\textbf{k}}\geq\varepsilon_{\text{F}},\\ 
  -\frac{\lambda_{\text{h}}}{\sqrt{2\pi\sigma^{2}}}e^{\frac{(\varepsilon_{n\textbf{k}}+\zeta)^{2}}{2\sigma^{2}}},
  & \varepsilon_{n\textbf{k}}<\varepsilon_{\text{F}}.\end{cases}\label{eq:initial-distribution}
\end{equation} 
Throughout this work, we choose a small energy smearing $\sigma=8.47$~meV and
small perturbation $\lambda_{\text{e}}=2.4 \times 10^{-3}$~\text{eV}. The parameter
$\lambda_{\text{h}}$ is selected such that the initially excited number of
electrons and holes is the same. Since the density of states (DOS) of graphene
is rather symmetric in the range of excitation energies
$-0.8~\text{eV}\leq\zeta\leq0.8~\text{eV}$ studied by us [see
  Fig.~\ref{fig:bandstruct}(c)], it turns out to be an excellent approximation
to set $\lambda=\lambda_{\text{e}}=\lambda_{\text{h}}$.

While we use $\lambda$ here as a free parameter to adjust the initial
occupation, it can be related to measurements through
$\lambda=4\pi^{2}\alpha\hbar^{2}\Phi v^2_{\text{F}}/\zeta^2$. In the
expression, $\pi\alpha$ is the linear absorption of graphene, $\Phi$ is the
pump-fluence and $v_{\text{F}}$ is the Fermi velocity of electrons in graphene
\citep{Maxim_Paper_PhysRevB.94.205306}.

We determine the time $\tau_{\text{th}}$, when hot carriers have relaxed through
the relation $P(\zeta,0,T)-P(\zeta,t,T)<P(\zeta,0,T)/\mathrm{e}$. In the
expression we have defined the population
\begin{equation}
  P(E,t,T)=\sum_{n\mathbf{k}}\delta(E-\varepsilon_{n\mathbf{k}})
  \begin{cases}\times[1-\tilde{f}_{n\mathbf{k}}(t,T)],&
    E<\varepsilon_{\text{F}},\\\times\tilde{f}_{n\mathbf{k}}(t,T),&
    E\geq\varepsilon_{\text{F}}.\end{cases} \label{eq:population}
\end{equation}
Our definition ensures that the population is symmetric with regard to
electrons and holes, as long as the DOS is symmetric.


\subsection{Analytical model}\label{subsec:method-analytical}

Before performing ab-initio calculations of charge carrier dynamics, we
estimate the photocarrier thermalization time of intrinsic graphene within an
explicitly solvable model. For simplicity we assume only optical phonon modes $p$ that
are dispersionless, i.e., exhibit the fixed energy $\hbar\omega_p$. For this
reason phonon wave vectors will be omitted. Furthermore, we consider only the
two linear electronic bands of the Dirac cone with
$\varepsilon_{n\mathbf{k}}=n\hbar v_{\text{F}} k$, $n=\pm$ and
$k=|\mathbf{k}|$. Additionally, we will suppress all time and temperature
arguments of the occupation functions in this subsection, while the tilde sign
will still be indicative of a time dependence of the electronic occupation
function.

The EP collisions in the given optical phonon mode $p$ are governed by the
following integral
\begin{eqnarray} I_{p}[\tilde{f}_{n\mathbf{k}}] & = & \sum\limits
  _{n'\mathbf{k}'}\left[\tilde{f}_{n'\mathbf{k}'}\left(1-\tilde{f}_{n\mathbf{k}}\right)W_{n'\mathbf{k}'\to
      n\mathbf{k}}\right.\nonumber \\ 
    & & \left.-\tilde{f}_{n\mathbf{k}}\left(1-\tilde{f}_{n'\mathbf{k}'}\right)W_{n\mathbf{k}\to
      n'\mathbf{k}'}\right], 
\end{eqnarray} 
where
$\tilde{f}_{n\mathbf{k}}=f_{\varepsilon_{n\mathbf{k}}}^{(0)}+\tilde{f}_{n\mathbf{k}}^{(1)}$
denotes the carrier occupation with the time-independent Fermi-Dirac
distribution $f_{\varepsilon_{n\mathbf{k}}}^{(0)}$ and the non-equilibrium
addition $\tilde{f}_{n\mathbf{k}}^{(1)}$, representing the second term in
  Eq.~(\ref{eq:Boltzmann-solution}). The transition probability is given by
Fermi's golden rule
\begin{eqnarray} 
  W_{n\mathbf{k}\to n'\mathbf{k}'} & = &
  \frac{2\pi}{\hbar}W_{p}\left[\left(N_{p}+1\right)\delta\left(\en-\enp-\hbar\omega_{p}\right)\right.\nonumber
    \\ & &
    \left.+N_{p}\delta\left(\en-\enp+\hbar\omega_{p}\right)\right]\label{w1}
\end{eqnarray}
for carriers outgoing from the state $|n\mathbf{k}\rangle$, and
\begin{eqnarray} W_{n'\mathbf{k}'\to n\mathbf{k}} & = &
  \frac{2\pi}{\hbar}W_{p}\left[\left(N_{p}+1\right)\delta\left(\enp-\en-\hbar\omega_{p}\right)\right.\nonumber
    \\ 
    & & \left.+N_{p}\delta\left(\enp-\en+\hbar\omega_{p}\right)\right]\label{w}
\end{eqnarray} 
for carriers incoming to the state $|n\mathbf{k}\rangle$. Making use of
  nearly dispersionless optical phonon modes, the EP interaction matrix
  element $W_p$ is assumed to be independent of momentum. The first term in
both Eqs.~(\ref{w1}) and (\ref{w}) corresponds to the phonon emission, while
the second one describes the phonon absorption. The phonons are treated as a
non-interacting gas, characterized by the Bose-Einstein distribution
$N_{p}= N_{\hbar\omega_{p}}(T)$. Due to the strong
carbon-carbon bonding in graphene the optical phonon energy is higher than
$100$~meV [see Fig.~\ref{fig:bandstruct}(c)] and, hence, we assume
$\hbar\omega_{p}\gg k_{\text{B}}T$ for typical temperatures or, in other
words, $N_{p}\ll1$. The collision integral can then be simplified to
\begin{widetext}
  \begin{equation}
    I_{p}[\tilde{f}_{n\mathbf{k}}] = \frac{2\pi}{\hbar}W_{p}\sum\limits
    _{n'\mathbf{k}'}\left[\tilde{f}_{n'\mathbf{k}'}\left(1-\tilde{f}_{n\mathbf{k}}\right)\delta\left(\enp-\en-\hbar\omega_{p}\right)
      -\tilde{f}_{n\mathbf{k}}\left(1-\tilde{f}_{n'\mathbf{k}'}\right)\delta\left(\en-\enp-\hbar\omega_{p}\right)\right].\label{Ieph1}
  \end{equation}
Let us now assume $\tilde{f}_{n\mathbf{k}}$ to be a function of $\en$ and
integrate in momentum space. Making use of the $\delta$-function and
$\varepsilon_{nk}=\varepsilon_{n\mathbf{k}}$, we obtain
\begin{equation}
I_{p}[\tilde{f}_{\varepsilon_{nk}}] =
\frac{W_{p}}{\hbar^{3}v^{2}_{\text{F}}}\left[|\varepsilon_{nk}+\hbar\omega_{p}|\tilde{f}_{\varepsilon_{nk}+\hbar\omega_{p}}\left(1-\tilde{f}_{\varepsilon_{nk}}\right)-|\varepsilon_{nk}-\hbar\omega_{p}|\tilde{f}_{\varepsilon_{nk}}\left(1-\tilde{f}_{\varepsilon_{nk}-\hbar\omega_{p}}\right)\right].\label{Ieph2}
\end{equation}
Finally, we employ a linear response approximation and the property of
intrinsic graphene $1-f_{\varepsilon_{nk}}^{(0)}=f_{-\varepsilon_{nk}}^{(0)}$ so that 
\begin{eqnarray} 
  \tilde{f}_{\varepsilon_{nk}+\hbar\omega_{p}}\left(1-\tilde{f}_{\varepsilon_{nk}}\right)&\approx&
  f_{\varepsilon_{nk}+\hbar\omega_{p}}^{(0)}f_{-\varepsilon_{nk}}^{(0)}-\tilde{f}_{\varepsilon_{nk}}^{(1)}f_{\varepsilon_{nk}+\hbar\omega_{p}}^{(0)}+\tilde{f}_{\varepsilon_{nk}+\hbar\omega_{p}}^{(1)}f_{-\varepsilon_{nk}}^{(0)}, \\
  \tilde{f}_{\varepsilon_{nk}}\left(1-\tilde{f}_{\varepsilon_{nk}-\hbar\omega_{p}}\right)&\approx& f_{-\varepsilon_{nk}+\hbar\omega_{p}}^{(0)}f_{\varepsilon_{nk}}^{(0)}+\tilde{f}_{\varepsilon_{nk}}^{(1)}f_{-\varepsilon_{nk}+\hbar\omega_{p}}^{(0)}-\tilde{f}_{\varepsilon_{nk}-\hbar\omega_{p}}^{(1)}f_{\varepsilon_{nk}}^{(0)}.
\end{eqnarray} 
Hence, Eq.~(\ref{Ieph2}) can be written as a sum of two terms
$I_{p}[\tilde{f}_{\varepsilon_{nk}}]=I_{p}[f_{\varepsilon_{nk}}^{(0)}]+\hat{I}_{p}[f_{\varepsilon_{nk}}^{(0)},\tilde{f}_{\varepsilon_{nk}}^{(1)}]$,
where
\begin{equation} 
 I_{p}[f_{\varepsilon_{nk}}^{(0)}]=\frac{W_{p}}{\hbar^{3}v^{2}_{\text{F}}}\left[|\varepsilon_{nk}+\hbar\omega_{p}|f_{\varepsilon_{nk}+\hbar\omega_{p}}^{(0)}f_{-\varepsilon_{nk}}^{(0)}-|\varepsilon_{nk}-\hbar\omega_{p}|f_{-\varepsilon_{nk}+\hbar\omega_{p}}^{(0)}f_{\varepsilon_{nk}}^{(0)}\right],\label{Ieph3gen}
\end{equation}
\begin{equation}
  \hat{I}_{p}[f_{\varepsilon_{nk}}^{(0)},\tilde{f}_{\varepsilon_{nk}}^{(1)}]=\frac{W_{p}}{\hbar^{3}v^{2}_{\text{F}}}\left[|\varepsilon_{nk}+\hbar\omega_{p}|\left(\tilde{f}_{\varepsilon_{nk}+\hbar\omega_{p}}^{(1)}f_{-\varepsilon_{nk}}^{(0)}-\tilde{f}_{\varepsilon_{nk}}^{(1)}f_{\varepsilon_{nk}+\hbar\omega_{p}}^{(0)}\right)-|\varepsilon_{nk}-\hbar\omega_{p}|\left(\tilde{f}_{\varepsilon_{nk}}^{(1)}f_{-\varepsilon_{nk}+\hbar\omega_{p}}^{(0)}-\tilde{f}_{\varepsilon_{nk}-\hbar\omega_{p}}^{(1)}f_{\varepsilon_{nk}}^{(0)}\right)\right].\label{Ieph4gen} 
\end{equation}
Eqs.~(\ref{Ieph3gen}) and (\ref{Ieph4gen}) are valid for any ratio between
$\varepsilon_{nk}$ and $\hbar\omega_{p}$ so that we can investigate the
thermalization behavior for photocarriers excited below and above the phonon
frequency. Note that only Eq.~(\ref{Ieph4gen}) is responsible for
thermalization, because Eq.~(\ref{Ieph3gen}) does not contain
$\tilde{f}_{\varepsilon_{nk}}^{(1)}$.

In what follows we consider the thermalization of electrons (i.e.,
$\varepsilon_k=\varepsilon_{+k}=\hbar v_{\text{F}}k$), as the thermalization
of holes is equivalent in the case of intrinsic graphene at not too high
excitation energies [see Fig.~\ref{fig:bandstruct}(c)]. Assuming the initial
non-equilibrium distribution to be $\delta$-shaped,
$f_{\varepsilon_k}^{(1)}\propto\delta\left(\varepsilon_k-\hbar\omega/2\right)$,
we find
\begin{equation} 
  I_{p}[\tilde{f}_{\varepsilon_k}]=\frac{\omega
    W_{p}}{2\hbar^{2}v^{2}_{\text{F}}}\left(\tilde{f}_{\varepsilon_k+\hbar\omega_{p}}^{(1)}f_{-\frac{\hbar\omega}{2}+\hbar\omega_{p}}^{(0)}+\tilde{f}_{\varepsilon_k-\hbar\omega_{p}}^{(1)}f_{\frac{\hbar\omega}{2}+\hbar\omega_{p}}^{(0)}\right)
  -\frac{W_{p}}{\hbar^{2}v^{2}_{\text{F}}}\tilde{f}_{\varepsilon_k}^{(1)}\left(\left|\frac{\omega}{2}+\omega_{p}\right|f_{\frac{\hbar\omega}{2}+\hbar\omega_{p}}^{(0)}+\left|\frac{\omega}{2}-\omega_{p}\right|f_{-\frac{\hbar\omega}{2}+\hbar\omega_{p}}^{(0)}\right). \label{eq:Iph5}
\end{equation} 
Eq.~(\ref{eq:Iph5}) contains cascade terms, generated each time, when a phonon
is emitted or absorbed \citep{PRB2011Malic}. We use the relaxation-time
approximation, i.e., we truncate the cascade to a single term proportional to
$\tilde{f}_{\varepsilon}^{(1)}$. This results in the thermalization time given
by
\begin{equation} 
  \frac{1}{\tau_{\mathrm{th}}}=\sum\limits_{p}\frac{W_{p}}{\hbar^{2}v^{2}_{\text{F}}}\left(\left|\frac{\omega}{2}+\omega_{p}\right|f_{\frac{\hbar\omega}{2}+\hbar\omega_{p}}^{(0)}+\left|\frac{\omega}{2}-\omega_{p}\right|f_{-\frac{\hbar\omega}{2}+\hbar\omega_{p}}^{(0)}\right).\label{tauth}
\end{equation} 
\end{widetext}  

This analytical model is of course not able to give quantitative predictions, but it
suggests that the thermalization time at $\omega_{p}\gg\omega/2$ is much
longer than at $\omega_{p}\ll\omega/2$. Indeed, in the latter limit we have
\begin{equation}
\frac{1}{\tau_{\mathrm{th}}}=\sum\limits _{p}\frac{\omega W_{p}}{2\hbar^{2}v^{2}_{\text{F}}}, \quad \omega_{p}   \ll \omega/2,
\label{eq:optical} 
\end{equation}
whereas in the former case the rate contains an exponentially small
multiplier, resulting in the following expression
\begin{equation} 
  \frac{1}{\tau_{\mathrm{th}}}=\sum\limits_{p}\frac{2\omega_{p}W_{p}}{\hbar^{2}v^{2}_{\text{F}}}{\mathrm{e}}^{-\frac{\hbar\omega_{p}}{k_{\text{B}}T}},
  \quad \omega_{p} \gg \omega/2.
  \label{eq:farinfra} 
\end{equation}
 
We will confirm the predictions of Eqs.~(\ref{eq:optical}) and
(\ref{eq:farinfra}) in the next section using the {\em ab-initio}
approach. Note, however, that while the approximation $\hbar\omega_{p}\gg
  k_{\text{B}}T$ or $N_{p}\ll1$, made for their derivation, is excellent for
  most temperatures studied, we will consider temperatures of up to 2000~K
  with our ab-initio approach, where this approximation becomes questionable.

\section{Results}\label{sec:results}

We will now use the ab-initio parameters for electrons, phonons and their
couplings, determined as described in
subsection~\ref{subsec:method-ab-initio}, and combine them with the Boltzmann
formalism of subsection~\ref{subsec:method-time-evolution} to study
photocarrier thermalization. At the end, we will compare to the results of the
analytical equations as derived in subsection~\ref{subsec:method-analytical}.

\begin{figure}[!tb]
  \begin{centering}
    \includegraphics[width=1.0\columnwidth]{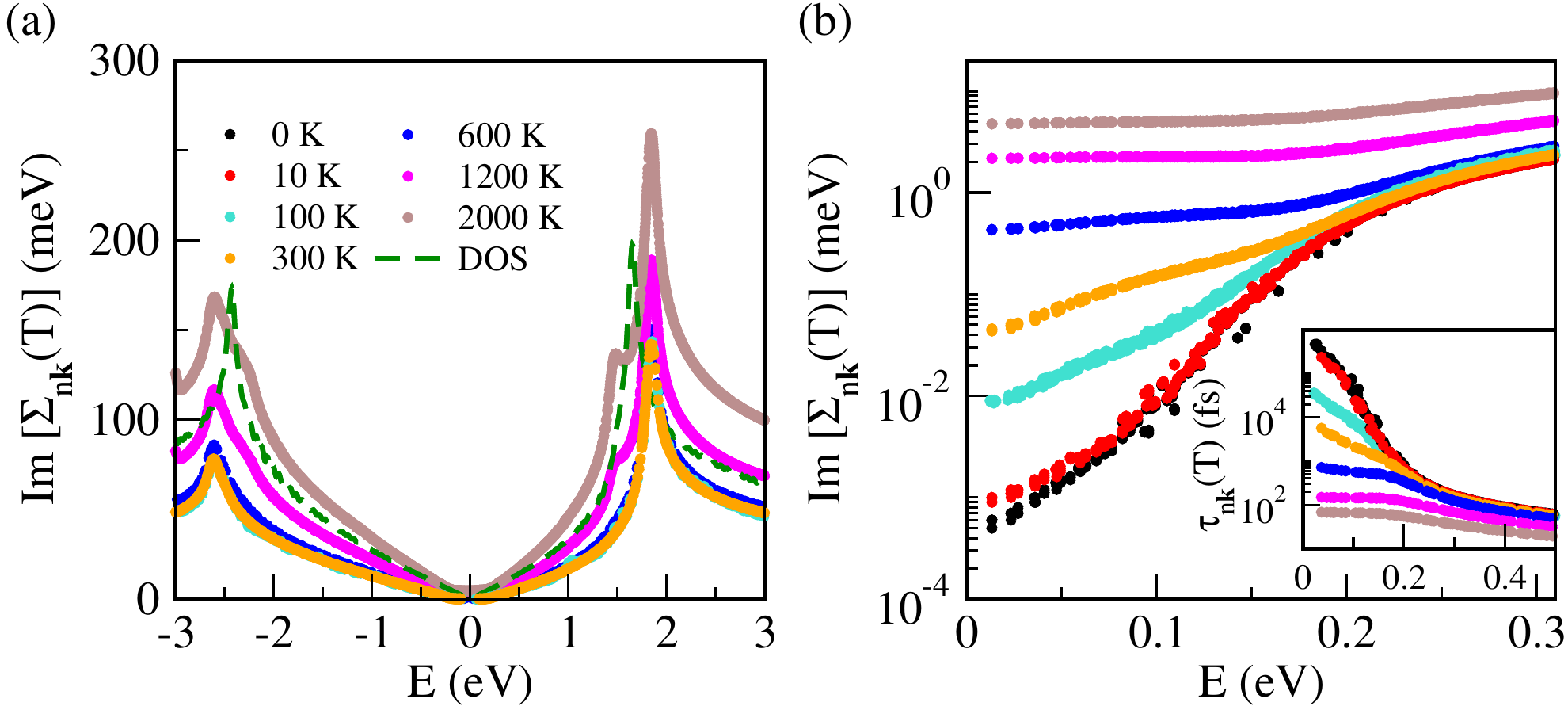} \par\end{centering}
    \centering{}\caption{(a) Imaginary part of the EP self-energy as a
      function of energy, evaluated at different temperatures, and the
      electronic DOS of graphene. (b) Zoom in on the energy and temperature
      dependence of $\text{Im}[\Sigma_{n\textbf{k}}(T)]$. We consider only
      positive energies close to the Dirac point. The inset represents the
      corresponding scattering} times.
    \label{fig:ImSigma-tau} 
\end{figure}

Since we determine scattering times $\tau_{n\mathbf{k}}$ of the Boltzmann
formalism [see Eq.~(\ref{eq:Boltzmann})] from the imaginary part of the
EP self-energy [see Eq.~(\ref{eq:self-energy})], we investigate
this quantity first. Fig.~\ref{fig:ImSigma-tau}(a) plots
$\text{Im}[\Sigma_{n\textbf{k}}(T)]$ as a function of energy for different
temperatures. For a given temperature it shows a pronounced energy dependence.
Increasing initially monotonically and rather symmetrically in the vicinity of
the Dirac point at $E=\varepsilon_{\text{F}}=0$, it follows the same behavior
as the electronic DOS [see Eqs.~(\ref{eq:self-energy})]. This results from the fact that the electronic DOS
represents the phase space for EP scattering events to take place. As can be
inferred from Fig.~\ref{fig:ImSigma-tau}(a) and \ref{fig:ImSigma-tau}(b),
$\text{Im}[\Sigma_{n\textbf{k}}(T)]$ is very sensitive to temperature close
to $E=0$. In contrast it shows a much weaker temperature dependence at
energies above around 200~meV, coinciding with the highest optical phonon
energies. Indeed, we see for low temperatures (0-300~K) that
Im$(\Sigma_{n\textbf{k}})$ increases roughly exponentially until the highest
optical phonon energy is reached, while the energy dependence is comparatively
weak for elevated temperatures (600-2000~K). The behavior shows that
scattering below the optical phonon threshold takes place rather inefficiently
via acoustical phonons. With increasing temperature there are more phonons
available for the carriers to interact with, leading to the increase of
$\text{Im}[\Sigma_{n\textbf{k}}(T)]$. Analogously, the available phase space
for optical phonon emission grows with increasing energy.

In the inset of Fig.~\ref{fig:ImSigma-tau}(b), we consider the scattering
times $\tau_{n\textbf{k}}(T)$, which are inversely proportional to the
self-energy [see Eq.~(\ref{eq:tau-nk})]. We observe that around the Dirac
point the scattering time becomes very sensitive to temperature and can be on
the order of a few picoseconds for low $T$. In contrast, at energies above
200~meV the scattering times exhibit only weak energy and temperature
dependencies. As argued before, this behavior can be rationalized by the fact
that for low $T$ at $E<200$~meV excited carriers can relax via acoustical
phonon scattering only, while they thermalize efficiently via optical phonons
above 200~meV.

The behavior of $\text{Im}[\Sigma_{n\textbf{k}}(T)]$ in
Fig.~\ref{fig:ImSigma-tau} can also be analyzed in terms of
Eq.~(\ref{eq:self-energy}). Lets consider low temperatures and electrons with
$\varepsilon_{n\mathbf{k}}\geq0$. In this case both
$f^{(0)}_{\varepsilon_{n\mathbf{k}}}(T)$ and
$N_{\hbar\omega_{p\mathbf{q}}}(T)$ are vanishingly small, and thus only the
second term of the Eq.~(\ref{eq:self-energy}) contributes. For this reason,
excited electrons relax via emission of phonons. But as temperature increases,
we get $0\leq f^{(0)}_{\varepsilon_{n\mathbf{k}}}(T)\leq1$ and
$N_{\hbar\omega_{p\mathbf{q}}}(T)>0$, and both terms in
Eq.~(\ref{eq:self-energy}) start contributing. For this reason
$\text{Im}[\Sigma_{n\textbf{k}}(T)]$ increases with increasing temperature in
Fig.~\ref{fig:ImSigma-tau} for $E>0$. An analogous argumentation can be
carried out for holes.

To simulate the temporal dynamics, we use Eq.~(\ref{eq:Boltzmann-solution}),
starting with the initial distribution of Eq.~(\ref{eq:initial-distribution})
at time $t=0$. Choosing the parameters $\lambda$ and $\sigma$ as described
above, we calculate time evolutions of occupations for different temperatures
$T$ and excitation energies $\zeta$. We are particularly interested in the
behavior of thermalization times for excitations below and above the optical
phonon threshold.

\begin{figure}[!tb]
  \begin{centering} 
    \includegraphics[width=1.0\columnwidth,height=0.5\textwidth,keepaspectratio]{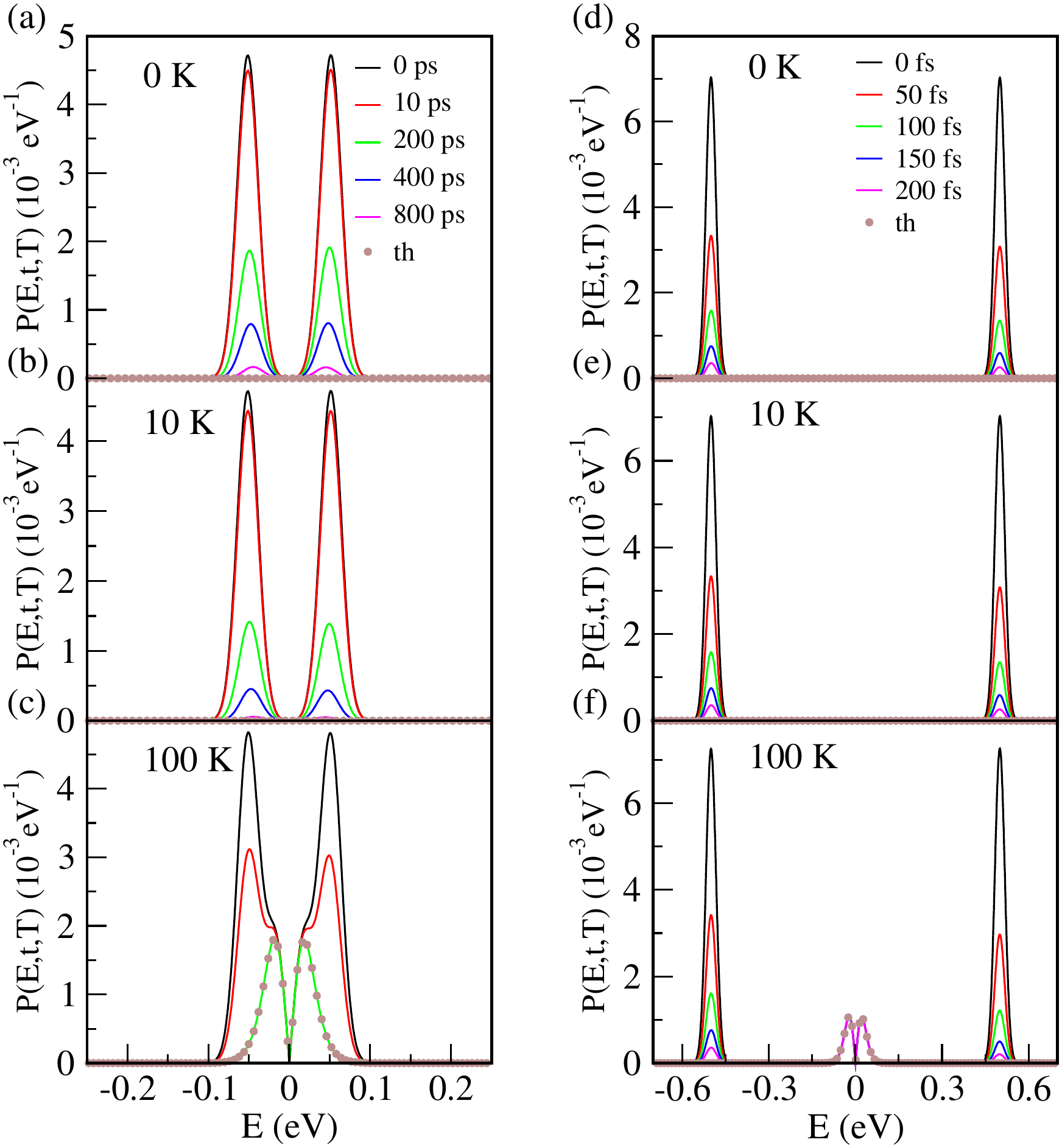} 
    \par\end{centering}
    \caption{Time-dependent thermalization of photocarriers at an excitation
      energy of (a)-(c) 0.05~eV and (d)-(f) 0.5~eV for different temperatures.}
    \label{fig:thermalization-tevol-lowT}
\end{figure}

Fig.~\ref{fig:thermalization-tevol-lowT} shows the hot carrier population
$P(E,t,T)$ [see Eq.~(\ref{eq:population})] for excitation energies
$\zeta=0.05, 0.5$~eV and temperatures $T=0,10,100$~K. Below the optical phonon
threshold for $\zeta=0.05$~eV in
Fig.~\ref{fig:thermalization-tevol-lowT}(a)-(c), thermalization of the hot
carriers takes place on the ps timescale via low-energy acoustical phonons. In
this excitation range the relaxation time decreases with increasing
temperature, because the background equilibrium electron distribution allows
excited carriers to scatter increasingly efficiently with the optical phonons
\citep{Acoustic_graphene_PhysRevLett.107.237401}. Our thermalization time
$\tau_{\text{th}}$ at $T=10$~K, as extracted from
Fig.~\ref{fig:thermalization-tevol-lowT}(b), is around 175~ps. This is lower
than the 300~ps reported in
Ref.~\citep{Acoustic_graphene_PhysRevLett.107.237401} for an excitation energy
of 51~meV on an epitaxially grown graphene sample containing around $\sim70$
layers and arranged over a SiC substrate. Above the optical phonon threshold,
our results in Fig.~\ref{fig:thermalization-tevol-lowT}(d)-(f) predict a weak
or almost no temperature dependence of the relaxation time. With
$\tau_{\text{th}}\approx60$~fs it takes a value of similar size as
  the photocarrier isotropization time from Ref.~\cite{Trushin_pseudospin},
originating from scattering by optical phonons. Our qualitative findings of a
strong temperature dependence of $\tau_{\text{th}}$ below the optical phonon
threshold and none above are consistent with the experimental observations in
Ref.~\citep{Acoustic_graphene_PhysRevLett.107.237401}. The plots in
Fig.~\ref{fig:thermalization-tevol-lowT} also demonstrate that the populations
of electrons and holes evolve with time quite symmetrically around the Dirac
point, confirming that the dynamics of holes are similar as those of
electrons.

Due to the extraordinarily high melting temperature of nearly $5000$~K predicted
theoretically for graphene \cite{melting_graphene}, we extend our analysis of
time evolutions to high temperatures $T=300,600,1200,2000$~K. 
We find carriers to relax at $T=300$ or $600$~K on a 100~fs time scale. At 1200~K this
reduces to around 34~fs and is even below 26~fs at 2000~K.

\begin{figure}[!tb] 
  \centering{}\includegraphics[width=1.0\columnwidth]{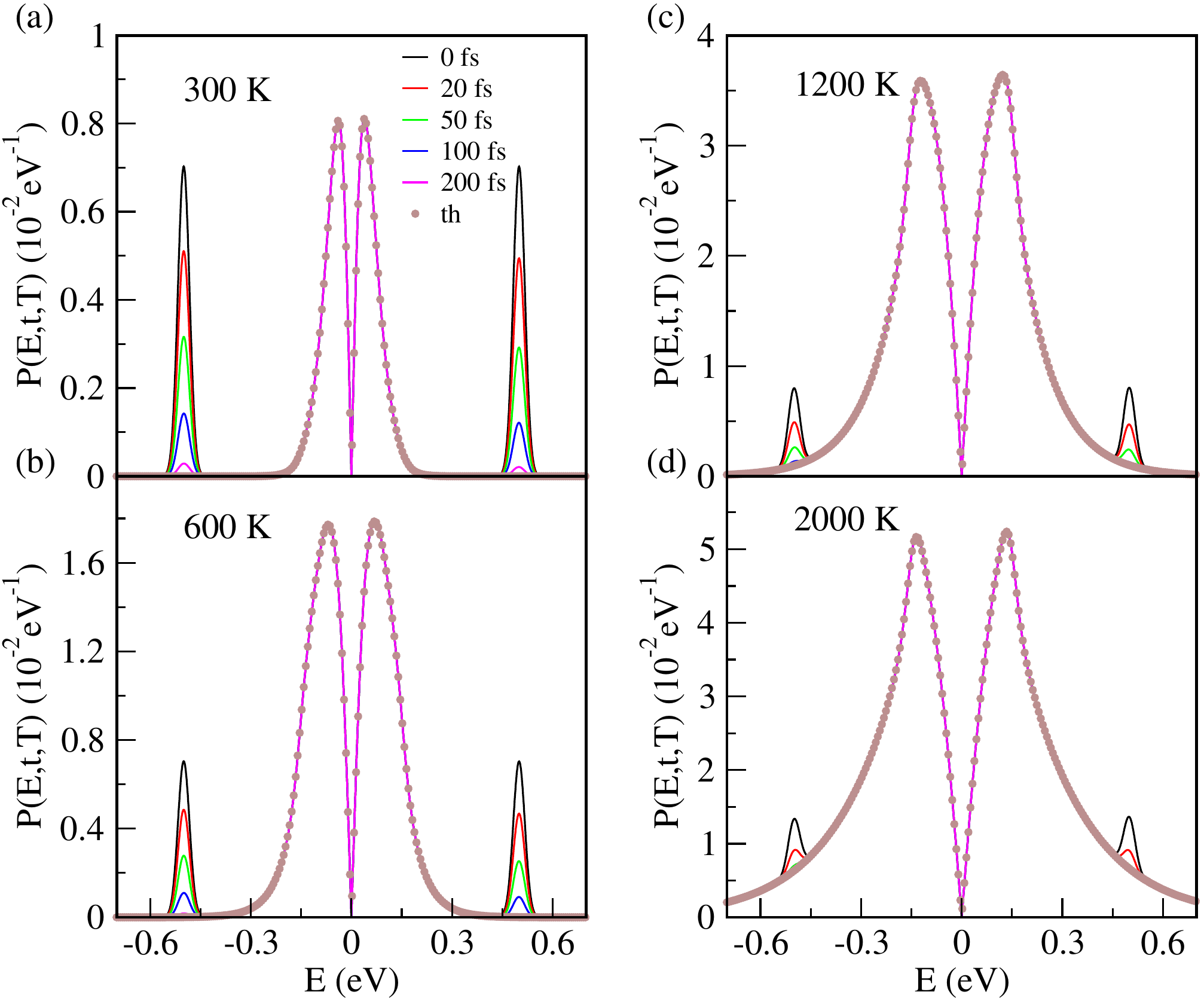} 
  \caption{Same as Fig.~\ref{fig:thermalization-tevol-lowT} at an excitation
    energy of 0.5~eV for elevated
    temperatures.} \label{fig:thermalization-tevol-highT}
\end{figure}

\begin{figure}[!tb]
  \centering{}\includegraphics[width=0.8\columnwidth]{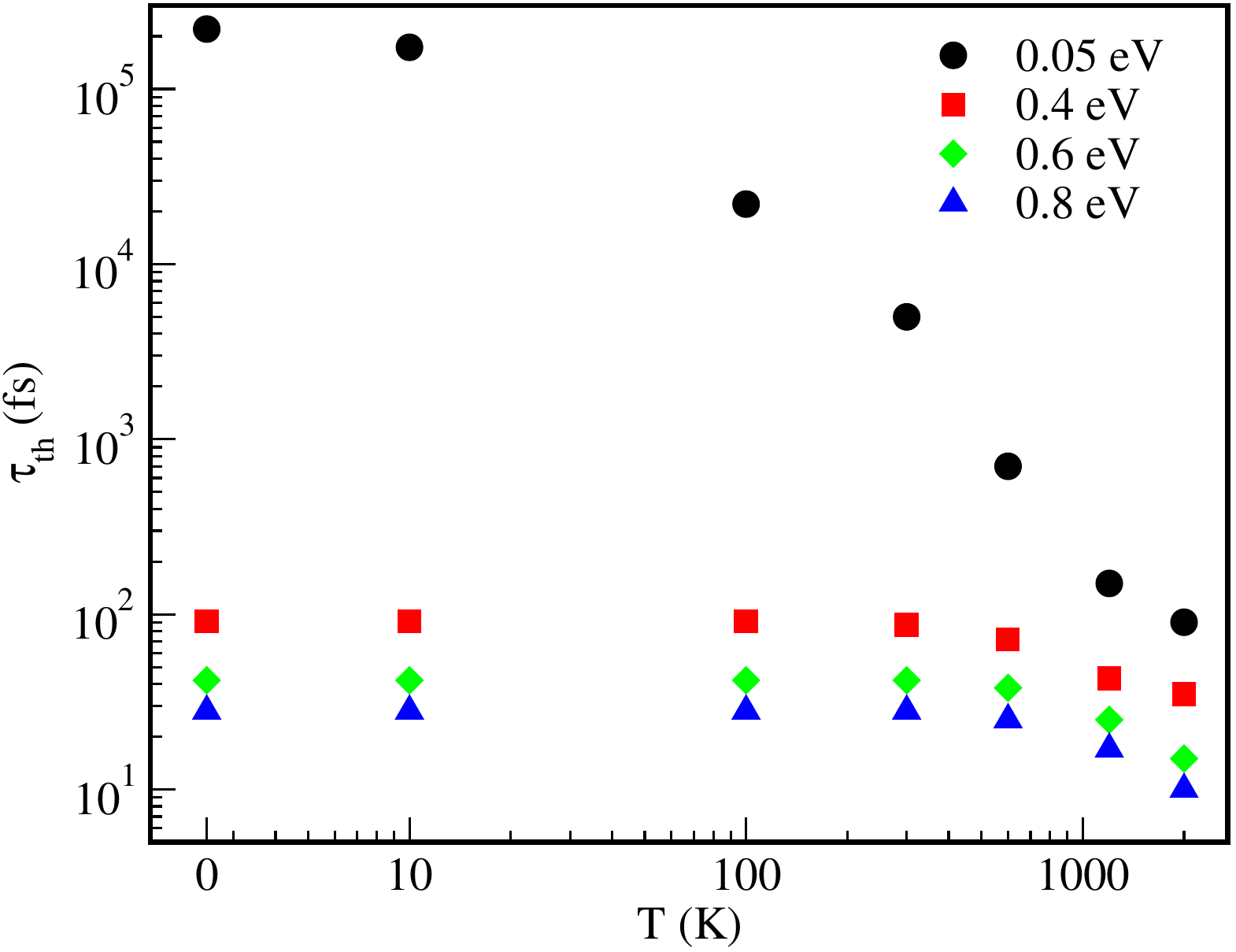}
  \caption{Thermalization time of the excited carriers, as determined with our
    ab-initio approach, as a function of temperature for different excitation
    energies.}
  \label{fig:t-th} 
\end{figure}

In Fig.~\ref{fig:t-th} we summarize the relaxation times $\tau_{\text{th}}$,
which we have extracted from our ab-initio modeling at different excitation
energies and temperatures. For $\zeta=0.05$~eV the thermalization time
decreases with increasing temperature from $T=0$ to $1200$~K by more than 3
orders of magnitude. In contrast, there is only little change in the
relaxation time with temperature for a fixed excitation with
$\zeta=0.4,0.6,0.8$~eV above the optical phonon threshold. A slight decrease
is seen at the temperatures, where thermal energies are similar to those of
optical phonon quanta, i.e., $k_\text{B}T\approx\hbar\omega_p$. In addition,
for a fixed temperature, relaxation times depend only little on $\zeta$, if
the excitation energy is above the optical phonon threshold. To summarize,
taking into account only EP scattering events, we thus observe intriguingly
that relaxation times in graphene can span an extraordinary range from 170~ps
down to 60~fs, if the temperature is varied and carriers are excited below the
optical phonon threshold.

Our ab-initio predictions can be qualitatively understood by using the concept
of a thermalization bottleneck in graphene. Thanks to the high optical phonon
energy quanta of about 200~meV [see Fig.~\ref{fig:bandstruct}(d)], the
low-energy (THz) electrons cannot relax as fast as the optically excited
photocarriers, because at low temperatures (i) the phonon absorption is a very
rare process and (ii) the phonon emission requires an empty electron state
below the Fermi level, but states below $\varepsilon_{\text{F}}$ are
almost fully occupied. The relevant thermalization times can be estimated
by using our analytical model. We assume an explicit form for the EP
interaction matrix element given by \citep{PRB2012low}
\begin{equation} 
  W_{p}=\frac{\hbar \Delta_{p}^{2}F_{p}}{2\rho\omega_{p}}, 
\end{equation} 
where $\Delta_{p}$ is the deformation potential for a mode $p$, $F_{p}$ is a
dimensionless geometric factor, and $\rho=7.6\times10^{-8}$~g/cm$^{2}$ is the
mass density. In what follows, we take into account the two most important
phonon modes \citep{PRB2012low}, $p=\Gamma,K$, where $F_{\Gamma}=1$,
$F_{K}=1/2$, $\hbar\omega_{\Gamma}=197$ meV, $\hbar\omega_{K}=157$ meV, and
$\Delta_{\Gamma}=\Delta_{0}$, $\Delta_{K}=\sqrt{2}\Delta_{0}$ with
$\Delta_{0}=11$~eV/\AA\ \citep{PRB2012low}. At $\omega\gg\omega_{p}$ the
thermalization time can be found from Eq.~(\ref{eq:optical}) as

\begin{equation}
  \tau_{\mathrm{th}}=\frac{4\omega_{0}}{\omega}\frac{\hbar v^{2}_{\text{F}}\rho}{
    \Delta_{0}^{2}},\quad\omega\gg\omega_{\Gamma,K}, 
\end{equation} 
where $1/\omega_{0}=1/\omega_{\Gamma}+1/\omega_{K}$. Assuming an excitation
energy of $\hbar\omega=1.55$~eV (i.e., a radiation wavelength of 800~nm), we
estimate $\tau_{\text{th}}\approx 58$~fs. In the opposite limit
$\omega\ll\omega_{p}$ we get from Eq.~(\ref{eq:farinfra})
\begin{equation} 
  \tau_{\mathrm{th}}=\frac{\hbar
    v^{2}_{\text{F}}\rho}{\Delta_{0}^{2}}\frac{1}{\exp(-\frac{\hbar\omega_{\Gamma}}{k_{\text{B}}T})+\exp(-\frac{\hbar\omega_{K}}{k_{\text{B}}T})},\quad\omega\ll\omega_{\Gamma,K}.
\end{equation}
Assuming the most relevant temperature of $300$~K, we estimate
$\tau_{\text{th}}\approx 92$~ps. 

Our considerations confirm that (i) the thermalization timescales differ at
$\omega\ll\omega_{p}$ and $\omega\gg\omega_{p}$ by three orders of magnitude
at room temperature, (ii) the photocarrier thermalization time strongly
depends on temperature at $\omega\ll\omega_{p}$, whereas at
$\omega\gg\omega_{p}$ it does not, (iii) in the former case, the
thermalization time decreases rapidly with increasing temperature. This is
exactly what we see in the summary of the relaxation times $\tau_{\text{th}}$ shown in
Fig.~\ref{fig:t-th}, as determined through our first principles approach.

\section{Summary and Outlook}\label{sec:summary-outlook}

In summary, we have studied the relaxation dynamics of hot carriers in
single-layer graphene near and away from the Dirac point subject to the EP
interaction. By determining electron and phonon dispersions as well as EP
couplings from DFT, our model based on the Boltzmann equation in the
relaxation-time approximation contains no free parameters and takes into
account contributions from all of the optical as well as acoustical branches
in the whole BZ. In excellent agreement with analytical predictions we find
that relaxation times computed with our ab-initio model are strongly enhanced,
if carriers are excited below the optical phonon energies. In addition, we
have shown that the carrier relaxation times depend strongly on temperature
for such low excitation energies, while being rather temperature-independent
for excitation energies above optical phonon energy quanta.

These effects could be employed to facilitate the photoexcited electron
transport from graphene to a semiconductor across a Schottky barrier
\cite{SciRep2014zhang,ACSNano2016defazio,massicotte2016photo}. Thanks to the
longer relaxation time at lower excitation energies, the photocarriers can
contribute to the interlayer transport before thermalization is completed,
thus improving the photoresponsivity \cite{trushin2017theory}. From the device
engineering point of view, the most important assumption made in this work is
the absence of a substrate. It might provide additional dielectric screening
and unintentional doping, which overall influence the electron-electron
scattering contribution neglected here. Moreover, the photocarriers might
experience interactions with remote polar surface phonons
\cite{PRB2012low}. Since the precise effects caused by a substrate strongly
depend on the chosen material and its interface properties, the model should be
tailored for each device to make quantitative predictions. Such a fine tuning
is out of scope here.

\section*{acknowledgment}

D.Y.\ and F.P.\ acknowledge financial support from the Carl Zeiss Foundation
as well as the Collaborative Research Center (SFB) 767 of the German Research
Foundation (DFG).  M.T.\ is supported by the Director's Senior Research
Fellowship from the Centre for Advanced 2D Materials at the National
University of Singapore (NRF Medium Sized Centre Programme
R-723-000-001-281). Part of the numerical modeling was performed using the
computational resources of the bwHPC program, namely the bwUniCluster and the
JUSTUS HPC facility.

\bibliography{hot-carrier-graphene} 
\end{document}